# Laser Cooling by Collisional Redistribution of Radiation

Ulrich Vogl and Martin Weitz

*Institut für Angewandte Physik der Universität Bonn, Wegelerstraße 8, 53115 Bonn, Germany*

The general idea that optical radiation may cool matter was put forward by Pringsheim already in 1929[1]. Doppler cooling of dilute atomic gases is an extremely successful application of this concept[2,3], and more recently anti-Stokes cooling in multilevel systems has been explored[4,5], culminating in the optical refrigeration of solids[6-9]. Collisional redistribution of radiation is a proposed different cooling mechanism that involves atomic two-level systems[10], though experimental investigations in gases with moderate density have so far not reached the cooling regime[11].

Here we experimentally demonstrate laser cooling of an atomic gas based on collisional redistribution of radiation, using rubidium atoms subject to 230 bar of argon buffer gas pressure. The frequent collisions in the ultradense gas transiently shift a far red detuned laser beam into resonance, while spontaneous decay occurs close to the unperturbed atomic resonance frequency. During each excitation cycle, a kinetic energy of order of the thermal energy $k_B T$ is extracted from the dense atomic sample. In a proof of principle experiment with a thermally non-isolated sample, we experimentally demonstrate relative cooling by 66 K. The cooled gas has a density of more than 10 orders of magnitude above the typical values in Doppler cooling experiments, and the cooling power reaches 87 mW. Future prospects of the demonstrated effect include **studies of supercooling beyond the homogeneous nucleation temperature[12,13] and optical chillers[9].**

Collisional redistribution is perhaps most widely known in the context of magneto-optical trapping of ultracold atoms, where this mechanism is a primary cause of trap loss processes[14]. In the long studied field of room-temperature interatomic collisions, redistribution of radiation is a natural consequence of line broadening effects due to collisionally aided excitation[15-18]. A remarkable issue is the extreme elasticity of collisions of excited state alkali atoms with atomic noble buffer gases[19]. We have recently shown that under ultrahigh buffer pressure gas conditions the frequent collisions allow for thermal equilibrium of coupled atom-light states[20], which holds prospects for a possible Bose-Einstein-like phase transition of atom-light quasiparticles[21-23].

The principle of the here reported cooling effect is illustrated in Fig. 1a. We use rubidium atoms in the presence of several hundred bar of buffer gas, for which the collisionally broadened linewidth approaches the thermal energy $k_B T$ in frequency units. In a quasimolecular binary collision picture, the energy splitting between the rubidium 5S ground state and the 5P excited state varies critically with the instantaneous spatial distance between a rubidium atom and a noble gas perturber. For a (red) laser detuning of a few nanometers, i.e. far outside the usual rubidium thin vapour absorption profile, excitation is enabled when a noble gas perturber approaches the rubidium atom and transiently shifts the D-line into resonance via the intermolecular potential, as an example of collisionally aided excitation[17]. The 27 ns natural lifetime of the rubidium 5P-state is orders of magnitudes longer than the ~ps duration of a collisional process. Provided that the collisions are elastic and the mean interatomic spacing is larger than the range of the potential, spontaneous decay of the excited state will most likely occur for larger distances from the perturber. The frequency of the emitted photon will correspondingly be closer to the unshifted atomic transition frequency. This completes the cooling cycle. For red laser detuning, in each excitation cycle kinetic energy is extracted from the sample. Moreover, entropy can be transferred to the photonic field, as the narrowband exciting laser radiation is converted into scattered fluorescence with frequency bandwidth distributed over the pressure broadened linewidth. The typical energy that can be removed from the sample in a cooling cycle is of order of the thermal energy $k_B T$, corresponding to $\sim h \cdot 10^{13}$ Hz for temperatures of order of the room temperature. This is a factor 20000 above the typical energy difference removed by a fluorescence photon in usual Doppler cooling of dilute atomic gases. We are aware that the described simple picture neglects multiparticle collisions.

To demonstrate the principle of laser cooling by collisional redistribution, we use a rubidium cell at initial temperature of 620 K (vapour limited number density $n_{Rb}$ $\sim 10^{16}$ cm$^{-3}$) and 230 bar (23 MPa) of argon buffer gas pressure ($n_{Ar} \sim 10^{21}$ cm$^{-3}$). Tunable laser radiation is sent into the steel high pressure gas cell through sapphire optical windows. Our setup is shown in Fig. 1b, and described in more detail in the methods section.

In initial experiments, the average optical frequency difference between incident and fluorescence light was measured. From energy conservation one finds that the expected laser cooling power in a sample is determined by

$$P_{cool,fl} = P_{opt} \, a(\nu) \, (\nu_{fl} - \nu) / \, \nu, \qquad (1)$$

where $P_{opt}$ denotes optical beam power, $a(\nu)$ the absorption probability for an incident optical frequency $\nu$, and $\nu_{fl}$ the average frequency of the scattered fluorescence. Fig. 2a shows typical results for the spectral distribution of the fluorescence for one red (red line) and one blue detuning (blue line) relatively to the rubidium D-lines. The corresponding incident optical frequencies are indicated by vertical thin lines respectively. In both cases the fluorescence spectrum is efficiently redistributed towards the line center. The average frequency shift here reaches 15-20 THz. The corresponding differential energy will cool (heat) the atomic sample for red (blue) laser detuning. The connected squares in Fig. 2b show the expected cooling power as a function of the laser frequency, as derived by recording different of such spectra and determining the cooling power following Eq. 1 from the average shift of the scattered fluorescence and the laser power absorbed in the rubidium sample. It is clearly seen that energy is extracted form the sample on the red side of the spectrum.

First evidence for an actual cooling of the gas was obtained with an IR-camera sensitive to blackbody radiation in the wavelength range 7.5-13 µm, which was directed to the outer surface of a sapphire window of our pressure cell. In this mid-infrared spectral region the sapphire window material is highly opaque, and in contrast to the gas sample we expect the window material to emit blackbody radiation characteristic for its temperature. Due to thermal transport in the sapphire material, a cooling of the dense gas will cause a temperature variation also at the outer surface of the cell window. For this measurement, the temperature change of the sapphire window following a 30 s long optical cooling period was determined. Fig. 3 shows a corresponding result recorded for a cooling laser frequency of 365 THz, corresponding to a detuning of 16 THz to the red of the rubidium D-lines centroid. The observed averaged temperature drop is 0.31(3) K near the beam center. This result for the cooling of the outer surface of the cell window is limited mainly by heating of the highly thermally conducting sapphire window ($\kappa_{sapphire} \cong 200$ W/cm·K) from contact with the metal body of the cell, which for each absorbed photon experiences an energy increase of the photon energy. The temperature increase of the metal cell body is also clearly visible in the figure, and attributed mainly to absorption of scattered fluorescence radiation. The observed temperature drop of the cell window is in agreement with the result of a simple heat transport model calculation. We interpret the observed change in blackbody radiation as clear qualitative evidence for laser cooling based on collisional redistribution.

In subsequent experiments, we have probed temperature information directly from inside the gas cell by means of thermal deflection spectroscopy[25-27]. A (non-resonant) helium-neon laser probe beam is sent collinearly and slightly displaced from the cooling beam through the high pressure cell, as indicated in Fig. 1b. Radial temperature variations induced by the cooling beam cause a gradient of the refractive index of the gas, and result in a prism-like deflection of the probe beam. The circles in Fig. 2b show measured deflection angles versus the cooling laser frequency. For red detuning the probe beam is deflected towards the cooling beam, which indicates cooling, while for blue detuning heating is observed. It is also visible that the cooling regime is only reached when the laser frequency is detuned some 5 THz further to the red of the value at which it just equals the average fluorescence frequency, i.e. at which the expected cooling power given by Eq. 1 reaches zero. This is attributed to residual heating, as could be due to a small quenching of excited state rubidium atoms in the ultrahigh pressure buffer gas environment[19]. Our data could be explained by an average decrease of the 27 ns natural lifetime of the rubidium 5P-state due to quenching in inelastic collisions by roughly 1 ns at 230 bar argon buffer gas pressure, which would be consistent with the directly measured value of the lifetime change of 0(±2) ns. The ratio of elastic to inelastic collisions of excited state rubidium atom with argon rare gas atoms here exceeds 20,000 to 1.

To acquire the spatial temperature profile induced by the cooling beam, the lateral offset between cooling and probe beam was scanned. The obtained deflection data is shown by the red dots in Fig. 4 for a cooling laser frequency of 365 THz and a relative absorption of 90% in the 1 cm long gas cell, which fits well to results derived using a theoretical heat transfer model (black line). The cooling and probe beam diameters used in this measurement are 1 mm. From numeric integration of the data the lateral temperature profile near the cell entrance is derived (see methods), and the result is shown by the solid blue line. The obtained temperature drop in the center of the cooling beam is 66(13) K. This value is much larger than the above given result for the temperature drop detected at the outside of the cell window, which is easily understood by the thermal conductivity of the argon gas being much below that of the sapphire window material, so that the cooling within the gas volume near the beam focal region can proceed further. Our model used to simulate heat transfer in the gas cell volume assumes that the cooling rate is proportional to the optical intensity of the Gaussian laser beam, i.e. we neglect the influence of radiation transport, which assumes that $\omega << \ell_{abs}$, where $\omega$ denotes is the cooling beam radius and $\ell_{abs}=1/\alpha$ the absorption length. In general, we expect the influence of radiation transport to be less severe than in the area of ultracold atomic gases[14], since the influence of light forces in the buffer gas sample is much smaller, but we are aware that future work should investigate this effect in more detail. The shown dashed blue line is the temperature profile obtained from the theoretical heat transfer model, from which a cooling power of 87(20)mW can be estimated, corresponding to a cooling efficiency of 3.6(0.8)%. These values are in reasonable agreement with the expected cooling power derived from the differential frequency shift of the fluorescence photons (see Fig. 2b and the corresponding discussion). Note that the cooling power is 4 - 5 orders of magnitude above that achieved in Doppler cooling experiments and also above results reported in the laser cooling of solids[7].

In conclusion, evidence for redistribution laser cooling of a gas has been obtained by (i) an observed blue shift of the scattered fluorescence (ii) thermographic imaging of the cell window and (iii) thermal deflection spectroscopy. The measured temperature drop is presently limited by thermal conductivity of the argon gas, and could be improved by either focusing the cooling beam tightly into a relatively short, optically dense gas cell to deposit the available cooling power into a smaller size gas volume or by the use of a thermally isolated all-sapphire cell.

**Methods**

**Setup and preparatory measurements.** The experiment is performed in a high pressure stainless steel cell with volume of a few $cm^3$ and sapphire optical windows filled with rubidium vapor and argon buffer gas (see Ref. 20). The optical cooling beam is derived from a continuous-wave titanium-sapphire laser of 3 W output power, whose output is directed into the pressure cell through the sapphire windows. To measure the rubidium upper state lifetime in the pressure broadened system, the laser radiation was periodically chopped with an acousto-optic modulator and the emitted fluorescence was recorded with a photomultiplier tube.

**Thermal deflection spectroscopy.** For measurement of cell inside temperatures, the cooling beam is periodically blocked with a mechanical shutter and the deflection of a probe beam is monitored with a split photodiode. Temperature variations induced by the cooling beam cause a local density change and a variation of the refractive index $n$, following $\frac{dn}{dT} \cong -\frac{n-1}{T}$. A collinear probe beam separated by a distance $r$ is deflected by an angle $\varphi = \frac{n-1}{T} \int_{0}^{L} dz \frac{dT(r,z)}{dr}$, where $L$ denotes the cell length. Assuming that the cooling beam removes energy with a rate following its intensity distribution $I(r,z) = I_0 \exp(-\alpha z - 2 r^2 / \omega^2)$, where $\omega$ denotes the beam radius, for a long cell the heat transfer along the radial direction will dominate and the temperature distribution can be determined from the beam deflection as

$$\Delta T(r,z) = \frac{T}{n-1} \cdot \frac{\alpha e^{-\alpha z}}{1-e^{-\alpha L}} \int_{r}^{\infty} \varphi(r')dr' . \qquad (2)$$

For the refractive index of the dense gas at the 632 nm probe beam wavelength (which is far detuned from the rubidium D-lines), a value $n \cong 1.057(6)$ for argon is used,[28] were the quoted uncertainty is predominantly determined by the accuracy of our cell pressure measurement.

For comparison, our deflection data was fitted assuming a temperature profile derived from a heat transport model[26-28]

$$\Delta T_{th}(r,z) = \frac{P_{cool,th}}{4\pi\kappa} \cdot e^{-\alpha z} \cdot \left[ Ei\left(\frac{-2r^2}{\omega^2 + 8Dt}\right) - Ei\left(\frac{-2r^2}{\omega^2}\right) \right] \quad , \tag{3}$$

where cooling power and thermal diffusivity $D$ of the gas were left as free fit parameters, $t$ denotes the exposure time and a value $\kappa$=0.034 W/mK for the thermal conductivity was used[29]. At our pressure and temperature values $D = \kappa / c\rho$ is not very well known, and the derived value $D$=0.16(5) mm²/s is in agreement with published results[29, 30].

**References**

1. P. Pringsheim, Zwei Bemerkungen über den Unterschied von Lumineszenz- und Temperaturstrahlung. *Zeitschrift für Physik* **57**, 739–746 (1929).

2. T. W. Hänsch & A. L. Schawlow, Cooling of gases by laser radiation. *Optics Commun.* **13**, 68–69 (1975).

3. S. Chu, C.N. Cohen-Tanoudji & W.D. Phillips, Nobel lectures in physics 1997, *Rev. Mod. Phys.* **70**, 685 (1998).

4. N. Djeu & W. T. Whitney, Laser cooling by spontaneous anti-stokes scattering. *Phys. Rev. Lett.* **46**, 236–239 (1981).

5. C. Zander & K. H. Drexhage, in *Advances in Photochemistry*, D.C. Neckers, D.H. Volman, and G. v. Bünau (eds.) (Wiley, New York, 1995).

6. R. I. Epstein, M. Buchwald, B. Edwards, T. Gosnell & C. Mungan, Observation of laser induced fluorescent cooling of a solid. *Nature* **377**, 500–503 (1995).

7. C. W. Hoyt, M. P. Hasselbeck, M. Sheik-Bahae, R. I. Epstein, S. Greenfield, J. Thiede, J. Distel, and J. Valencia, Advances in laser cooling of thulium-doped glass. *J. Opt. Soc. Am. B* **20**, 1066-1074  (2003).

8. J. Thiede, J. Distel, S. R. Greenfield & R. I. Epstein, Cooling to 208 K by optical refrigeration. *Appl. Phys. Lett.* **86**, 154107 (2005).

9. M. Sheik-Bahae & R. I. Epstein, Optical refrigeration. *Nature Photonics* **12**, 693–699 (2007).

10. P. R. Berman & S. Stenholm, Heating or cooling using collisionally aided fluorescence. *Opt. Commun.* **24**, 155–157 (1978).

11. E. Giacobino, M. Tawil, P. R. Berman, O. Redi & H. H. Stroke, Production of "hot" excited-state atoms in collisionally aided radiative transitions.  *Phys. Rev. A* **28**, 2555-2557 (1983).

12. P. G. Debenedetti & H. E. Stanley, Supercooled and Glassy Water. *Phys. Today* **56**, 40-46 (2003).

13. T. Koop, B. Luo, A. Tsias & T. Peter. Water activity as the determinant for homogeneous ice nucleation in aqueous solutions. *Nature* **406**, 611-614 (2000).

14. C. S. Adams & E. Riis, Laser cooling and trapping of neutral atoms. *Progress in Quantum Electronics* **21**, 1–79 (1997).

15. C. Cohen-Tannoudji, J. Dupont-Roc & G. Grynberg, Atom-Photon Interactions – Basic Processes and Applications (Wiley, New York, 1992).

16. F. Schuller & W. Behmenburg, Perturbation of spectral lines by atomic interactions. *Phys. Rep.* **12**, 273–334 (1974).

17. S. Yeh & P. R. Berman, Theory of collisionally aided radiative excitation. *Phys. Rev. A* **19**, 1106–1116 (1979).

18. R. E. M. Hedges, D. L. Drummond & A. Gallagher, Extreme-wing line broadening and Cs-inert-gas potentials. *Phys. Rev. A* **6**, 1519–1544 (1972).

19. E. Speller, B. Staudenmayer & V. Kempter, Quenching cross sections for alkali-inert gas collisions. Z. Phys. A **291**, 311–318 (1979).

20. U. Vogl & M. Weitz, Spectroscopy of atomic rubidium at 500-bar buffer gas pressure: Approaching the thermal equilibrium of dressed atom-light states, Phys. Rev. A **78**, 011401 (2008).

21. P. R. Eastham & P. B. Littlewood, Bose condensation of cavity polaritons beyond the linear regime: The thermal equilibrium of a model microcavity, *Phys. Rev. B* **64**, 235101 (2001).

22. H. Deng, G. Weihs, C. Santori, J. Bloch & Y. Yamamoto, Condensation of semiconductor microcavity exciton polaritons, *Science* **298**, 199-2002 (2002).

23. C. Bolkart, R. Weiss, D. Rostohar & M. Weitz, Coherent and BCS-Type Quantum States of Dark Polaritons, *Las. Phys.* **15**, 3-6 (2005).

24. J. Pascale and J. Vandeplanque, Excited molecular terms of the alkali-rare gas atom pairs, *J. Chem. Phys.* **60**, 2278 (1974).

25. J. Whinnery, Laser measurement of optical absorption in liquids, *Accounts of Chemical Research* **7**, 225 (1974).

26. W. B. Jackson, N. M. Amer, A. C. Boccara & D. Fournier, Photothermal deflection spectroscopy and detection. *Appl. Opt.* **20**, 1333–1344 (1981).

27. J. D. Spear, R. E. Russo & R. J. Silva, Collinear photothermal deflection spectroscopy with light scattering samples. *Appl. Opt.* **29**, 4225–4234 (1990).

28. M. Born and E. Wolf, Principles of Optics (7th ed.), Pergamon Press (1999).

29. B. M. Rosenbaum, S. Oshen & G. Thodos, Thermal conductivity of argon in the dense gaseous and liquid regions. *J. Chem. Phys.* **44**, 2831–2838 (1966).

30. J.-M. P. Tournier & M. S. El-Genk, Properties of noble gases and binary mixtures for closed brayton cycle applications. *Energy Conversion and Management* **49**, 469–492 (2008).

**Acknowledgements**

We thank J. Nipper for experimental contributions during the early phase of this project. Financial support from the Deutsche Forschungsgemeinschaft within the focused research unit FOR557 is acknowledged.

**Competing financial interests**

The authors declare that they have no competing financial interests.

**Figure Captions:**

Figure 1: **Cooling principle and setup. a,** The black solid lines schematically indicate rubidium atomic energy levels versus the separation from a noble gas perturber (see Ref. 24 for more realistic, modelled alkali-noble gas potentials). Far red detuned laser radiation can be absorbed when a perturber atom approaches the rubidium atom. Spontaneous decay occurs mostly at larger distances of the perturber, so that the emitted photons have higher frequency than the incident radiation and the atomic ensemble looses energy. **b**, Experimental setup for demonstration of laser cooling of an ultradense gas by collisional redistribution of fluorescence.

Figure 2: **Experimental spectra. a**, Fluorescence spectra of the rubidium D-lines at 230 bar argon buffer gas pressure for an incident laser frequency of 360 THz (red line) and 404 THz (blue line), corresponding to a red and a blue detuning respectively. The incident laser frequencies are indicated as vertical thin lines. In both cases, the fluorescence frequency is efficiently redistributed towards the D-lines center, as indicated by the shown arrows. The visible small spikes are attributed mainly to stray excitation light. The rubidium resonance frequencies are 377 THz and 384 THz for the D1- and D2-line respectively. **b**, The connected blue squares show the expected cooling power versus laser frequency, as derived from the average fluorescence frequency shift. The red

circles (fitted with a spline function) give measured data for the angle deflection of a probe beam.

Figure 3: **Thermographic image.** An IR-camera is directed to the sapphire entrance window of the high pressure buffer gas cell. Shown is the (averaged) colour-coded temperature change after a 30 seconds long pulse of the cooling beam. The blue region in the center is the sapphire window, cooled by the gas sample from behind. The surrounding metal flange appears red and yellow, due to heating from absorption of scattered fluorescence radiation. To avoid a possible residual influence of cooling laser radiation onto the IR-camera signal, the camera was enabled only after the end of a cooling laser pulse. The used diameter of the cooling beam was 3 mm.

Figure 4:  **Measurement of cell inside temperature**. The red dots show results for the probe beam angle deflection versus the lateral offset between probe and cooling beams, and the black line gives a fit to a theoretical model. The deflection gives a measure for the radial temperature gradient induced by the cooling beam (see methods). The solid blue line shows the corresponding temperature profile near the cell entrance. The determined temperature drop in the beam center is 66(13) K. The quoted error bar is determined mainly by uncertainties in the optical beam diameter and the beam geometry, the measured absorption coefficient for the cooling radiation and the buffer gas

pressure measurement. The dashed blue line follows a heat transport model

(Eq. 3).

**a**

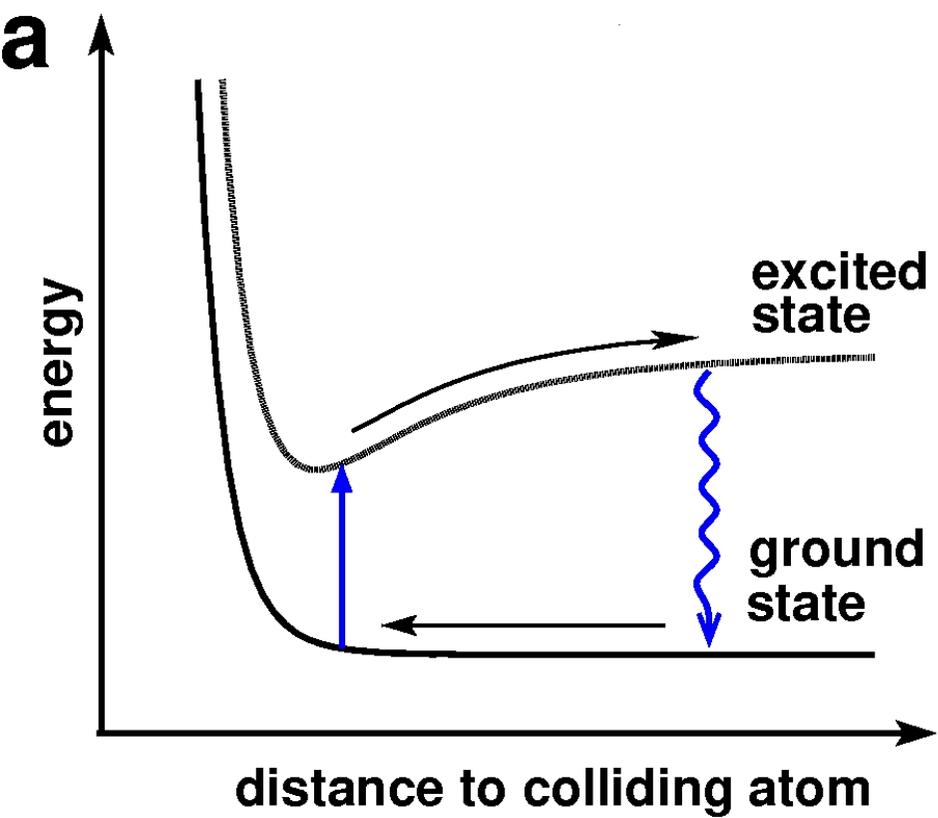

energy

excited
state

ground
state

distance to colliding atom

**b**

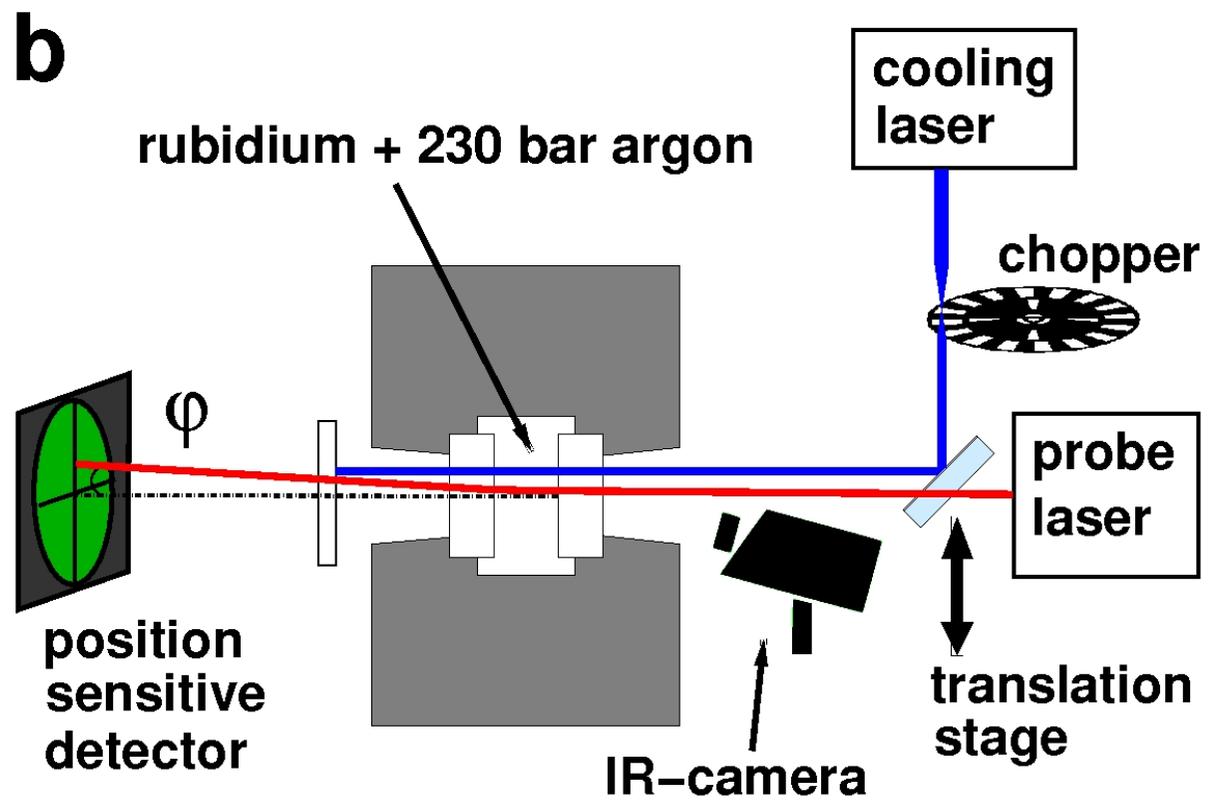

rubidium + 230 bar argon

cooling
laser

chopper

φ

probe
laser

position
sensitive
detector

IR−camera

translation
stage

**a**

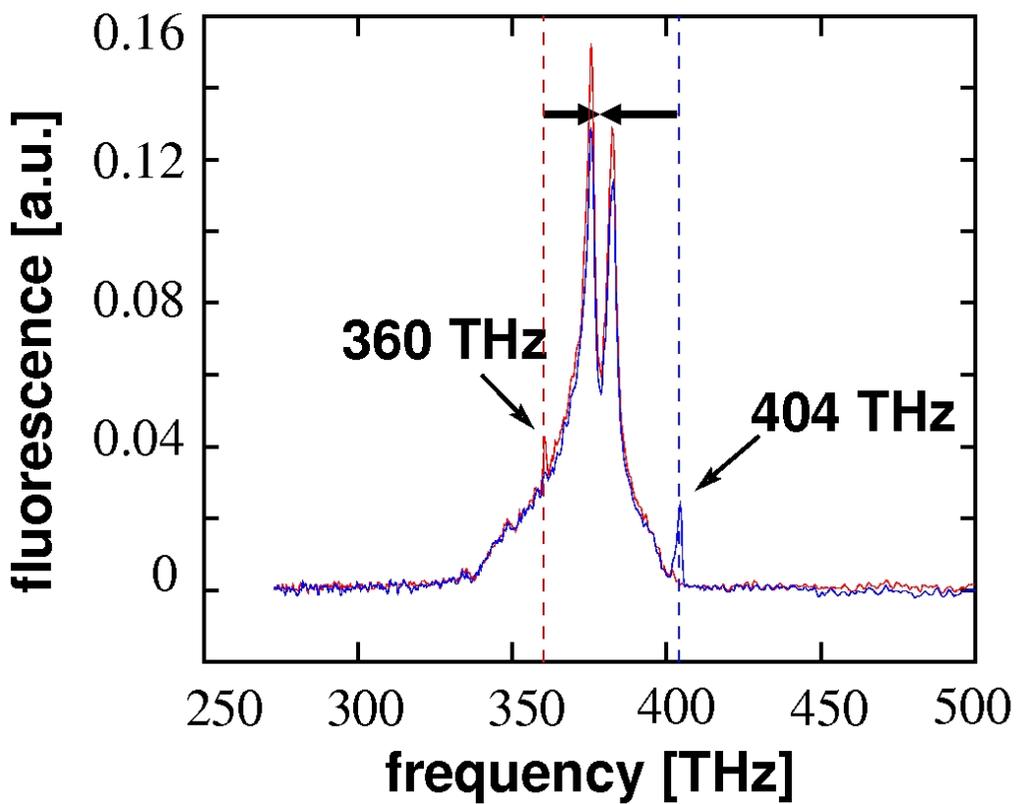

**b**

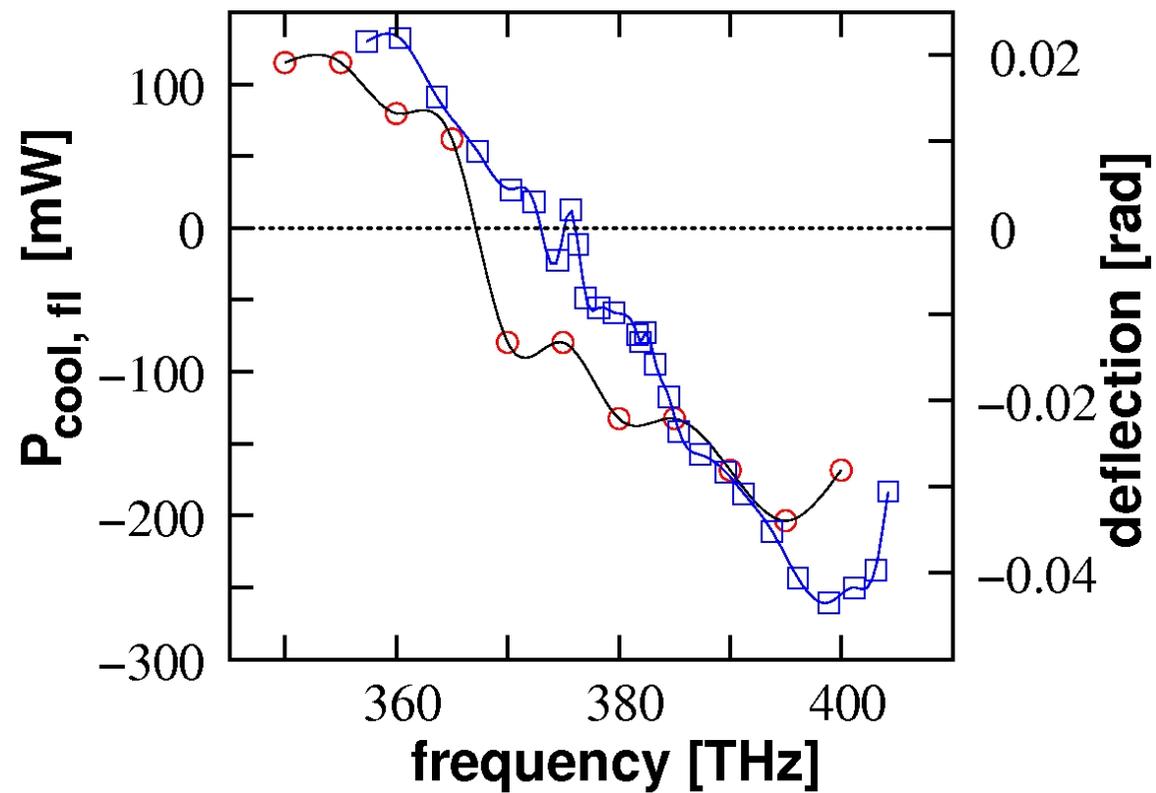

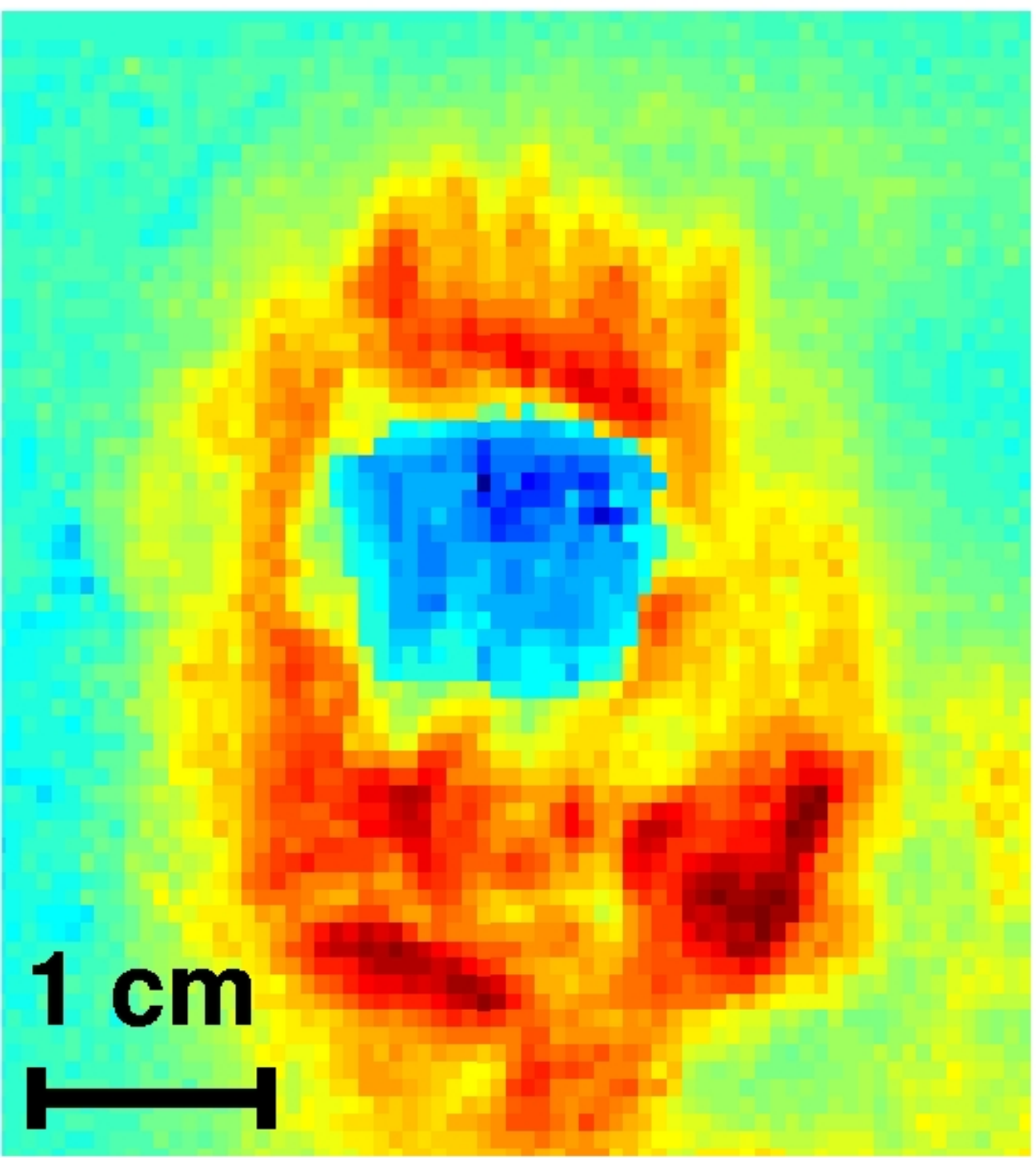
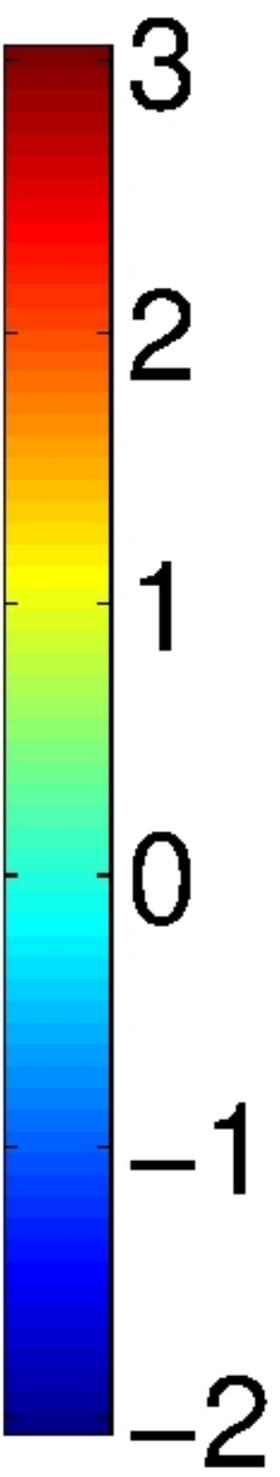

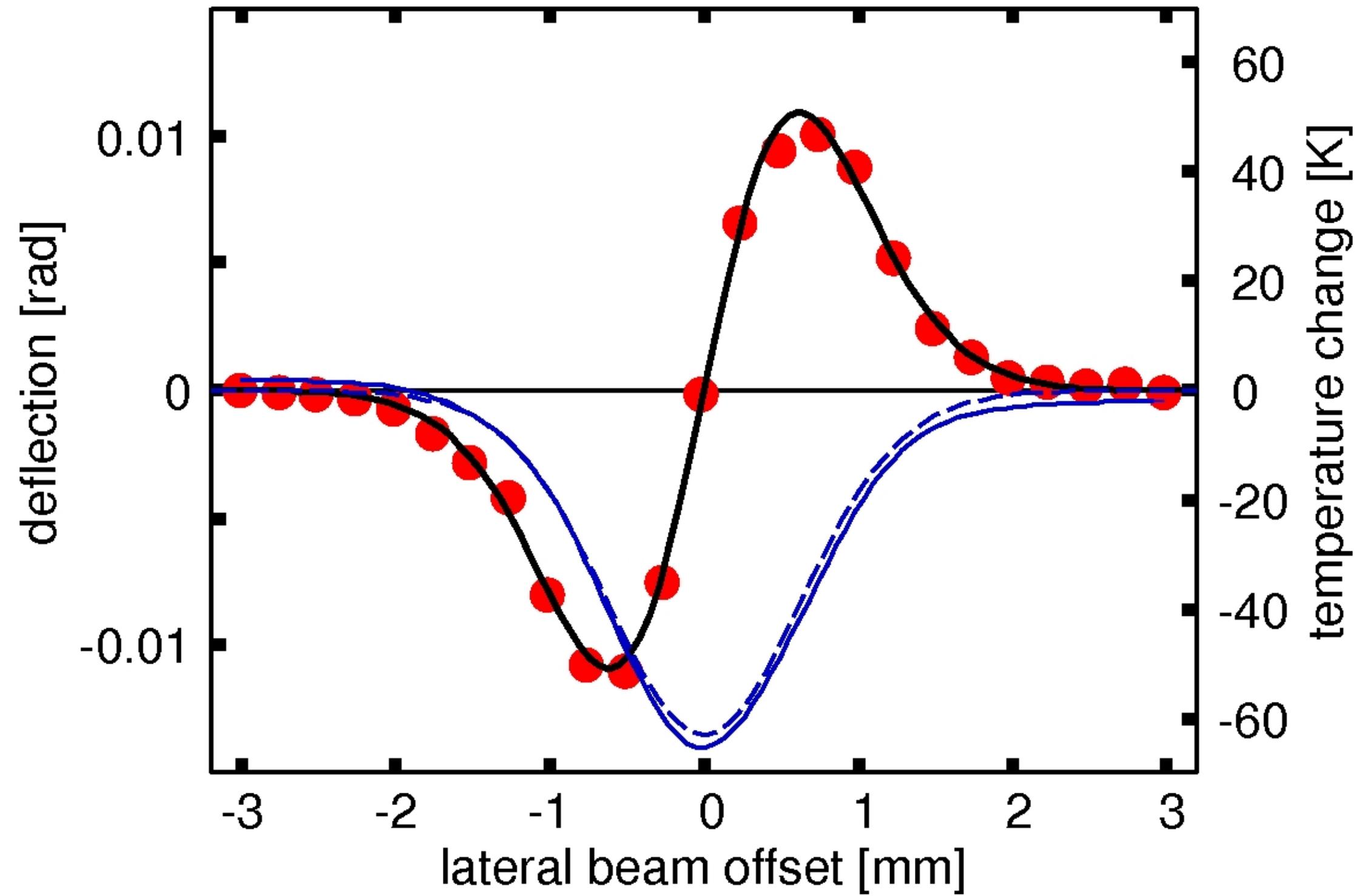